\documentclass[aps,prl,floatfix,superscriptaddress,titlepage,twocolumn,amsmath,amssymb,longbibliography]{revtex4-1}

\usepackage{graphicx}
\usepackage{bm}
\usepackage{braket}
\usepackage{times}
\expandafter\let\csname equation*\endcsname\relax
\expandafter\let\csname endequation*\endcsname\relax
\usepackage{amsmath}
\usepackage{amssymb}
\usepackage{amstext}

\usepackage{color}
\definecolor{darkgreen}{rgb}{0.2,0.8,0.2}
\definecolor{darkred}{rgb}{0.8,0.2,0.2}
\definecolor{darkblue}{rgb}{0.2,0.2,0.8}
\usepackage[pdftex,
        colorlinks=true,
        linkcolor=darkblue,
        citecolor=darkblue,
        filecolor=black,
        urlcolor=darkblue,
        bookmarks=false,
        bookmarksopen=false,
        bookmarksopenlevel=3,
        plainpages=false,
        pdfpagelabels=true,
        verbose=true]{hyperref}
\usepackage[all]{hypcap}

\newcommand{\up}{\uparrow}
\newcommand{\down}{\downarrow}
\renewcommand{\:}{\,$$=$$\,}
\newcommand{\ER}{E_R}
\newcommand{\Vdip}{V_\text{dip}}
\newcommand{\Vd}{\epsilon_\text{dip}} % relative strength
\newcommand{\Vh}{V_\text{hc}}
\newcommand{\Vz}{V_\text{1D}}
\newcommand{\Supp}{Supplemental Material \cite{Supp}}

\newcommand{\wdip}{w_\text{dip}}
\newcommand{\e}{\mathrm{e}}
\newcommand{\q}{\mathbf{q}}
\newcommand{\I}{\mathrm{i}}
\newcommand{\Ns}{S}

\newcommand{\Fig}[1]{Fig.~\ref{#1}}

\begin{document}
 \topmargin-1.5cm
\textheight22cm
\title{~\\Emulating Molecular Orbitals and Electronic Dynamics with Ultracold Atoms}

\newcommand{\ILP}{\affiliation{Institut f\"ur Laserphysik, Universit\"at Hamburg, Luruper Chaussee 149, 22761 Hamburg, Germany}}

\author{Dirk-S\"oren L\"uhmann}\ILP
\author{Christof Weitenberg}\ILP
\author{Klaus Sengstock}\ILP

\begin{abstract}
In recent years, ultracold atoms in optical lattices have proven their great value as quantum simulators for studying strongly correlated phases and complex phenomena in solid-state systems. Here we reveal their potential as quantum simulators for molecular physics and propose a technique to image the three-dimensional molecular orbitals with high resolution. The outstanding tunability of ultracold atoms in terms of potential and interaction offer fully adjustable model systems for gaining deep insight into the electronic structure of molecules. We study the orbitals of an artificial benzene molecule and discuss the effect of tunable interactions in its conjugated $\pi$ electron system with special regard to localization and spin order. The dynamical time scales of ultracold atom simulators are on the order of milliseconds, which allows for the time-resolved monitoring of a broad range of dynamical processes. As an example, we compute the hole dynamics in the conjugated $\pi$ system of the artificial benzene molecule. 
\end{abstract}

\maketitle

The structure of molecules is usually determined by x-ray or electron diffraction. Current advances with femtosecond pulses allow for the time-resolved observation of the atomic positions \cite{Miller2014}. In general, orbital wave functions are much harder to image. For simple molecules, the reconstruction of the HOMO has been achieved recently by electron momentum spectroscopy \cite{Brion2001}, by using laser-induced electron diffraction \cite{Meckel2008}, and by higher-order harmonic generation \cite{Itatani2004,Haessler2010}. For hydrogen atoms the nodal lines of Stark states were resolved via photoionization microscopy \cite{Stodolna2013}. State-of-the-art scanning probe microscopy allows resolving the HOMO as well as identifying the chemical bonds \cite{Gross2011} for benzene-based compounds attached to surfaces. 

Experimental access to the electronic structure of molecules and their dynamics is essential because even for relatively small molecules the full many-particle problem is not computable using classical computers. This has brought forward the idea of quantum computation for molecules \cite{Aspuru-Guzik2005,Kassal2011}. The quantum computation of a hydrogen molecule in a minimal basis has already succeeded \cite{Lanyon2010,Du2010}, but estimates for more complex molecules are not promising. While the number of required qubits appears manageable, the estimated number of quantum gate operations ($10^{18}$ for Fe$_2$S$_2$) is many orders of magnitude larger than currently available \cite{Wecker2014}. 

In the past decade, ultracold atoms in optical lattices have been established as quantum simulators for condensed matter systems \cite{Bloch2008,Lewenstein2007}. In addition, ultracold atoms were proposed as simulators for different systems such as neutron stars \cite{Gezerlis2008}, black holes \cite{Garay2000,Leonhardt2000}, quarks \cite{Rapp2007}, or atoms \cite{Sala2013}. Recent experimental developments include prospects such as the single-lattice-site imaging and addressing \cite{Bakr2009,Sherson2010} as well as the deterministic preparation and detection of few-atom samples \cite{Serwane2011,Wenz2013}. 

Here we show that ultracold atoms can be employed as a quantum simulator for molecules using existing experimental techniques. In this setting, the ultracold atoms mimic the electrons in a molecule, whereas the optical trapping potential takes the role of the nuclei. We demonstrate that ultracold atoms can serve as a tunable model system allowing the investigation of open questions in molecular physics. In contrast to the quantum computation approach aiming at the exact calculation of molecular energies, the present quantum emulation approach uses model systems to investigate specific interaction effects and dynamical processes in moleculelike systems. As a concrete example, we focus on a model system for benzene and compute its molecular orbitals for vanishing interaction. For the nonsolvable interacting problem, ultracold atoms can serve as a quantum simulator for static and dynamical electronic properties in molecules. We demonstrate numerically that the conjugated $\pi$ electron system shows strong-correlation effects such as localization and spin order, even when neglecting the interaction with inner particles. On the basis of this subsystem, we also reveal that ultracold atom simulators promise unique insight into electronic femtosecond dynamics. We show that momentum mapping in combination with phase retrieval allows the imaging of molecular orbitals with 1-2 orders of magnitude better than the intramolecular distances. We explain how to use the outstanding tunability of interaction and potential for studying electronic interactions and the dynamics in artificial molecules.

%%%%%%%%%%%%%%%%%%%%%%%%%%%%%%
\begin{figure*}
\centering \includegraphics[width=\linewidth]{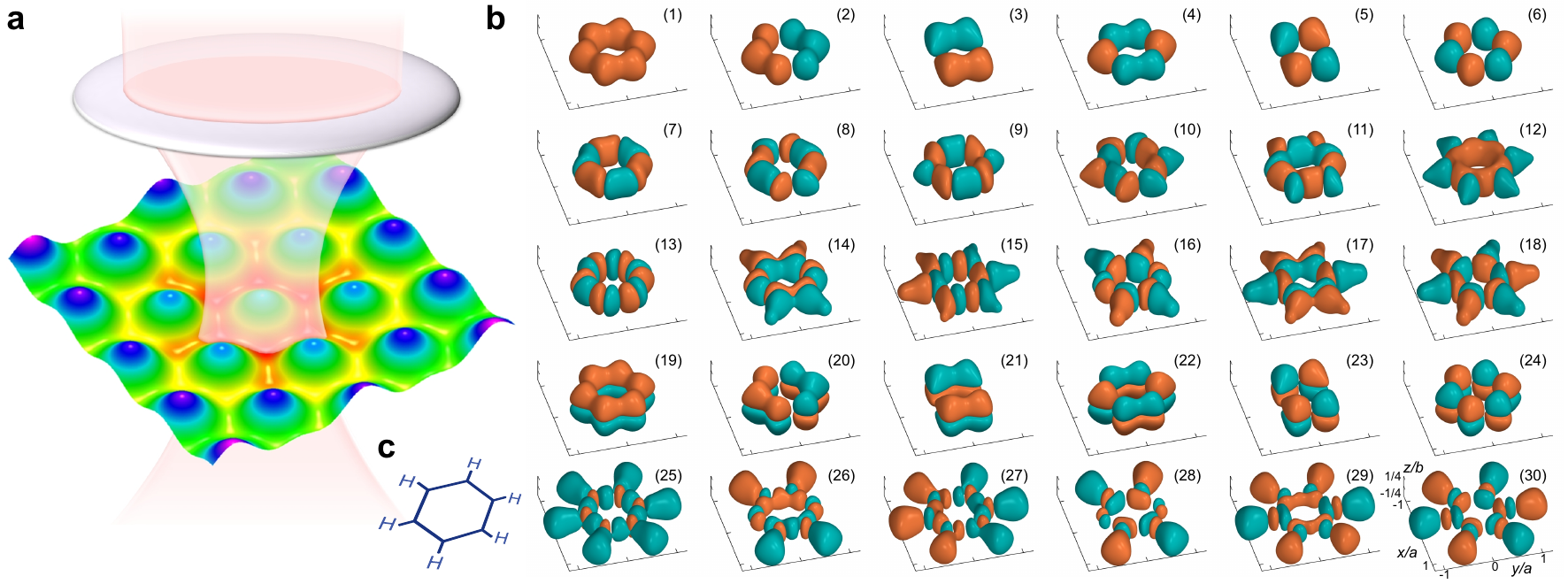}
\caption{Molecular orbitals of artificial benzene. (a) Illustration of a honeycomb optical lattice superposed with a dipole trap. The hexagonal ring and the adjacent sites form a trapping potential for an artificial, benzenelike molecule. (b) Calculated molecular single-particle orbitals of the artificial benzene molecule with low-lying $s$ orbitals (1-6) on the inner "C" ring, hybridized $sp^2$ orbitals (7-18,25-30) including the adjacent "H" atoms, and $p_z$ orbitals (19-24) forming a conjugated $\pi$ system. The orbitals are plotted for the lattice depths $V_0=11\ER$ and $V_{0z}=35\ER$ in units of the recoil energy $\ER$ of the lattice wavelength. The isosurfaces depict the orbital wave functions at $0.4/a\sqrt{b}$ (orange) and $-0.4/a\sqrt{b}$ (green) with the lattice constant $a$ of the honeycomb and $b$ of the orthogonal lattice. (c) Chemical structure of the benzene molecule C$_6$H$_6$.} \label{Fig1}
\end{figure*}
%%%%%%%%%%%%%%%%%%%%%%%%%%%%%%

\section{I. Creating artificial molecules}
As an example, we discuss how to create an artificial benzene molecule and compute the orbital wave functions. In real benzene, the molecular structure is formed by a ring of six carbon atoms and six hydrogen atoms (\Fig{Fig1}c). Thereby, the molecular symmetry is essential for the formation of molecular orbitals, where benzene belongs to the point group $D_\text{6h}$. In our case, the idea is to simulate the electrons in a molecular structure via ultracold atoms in a tailored optical dipole potential that mimics the  Coulomb interaction with the nuclei. The optical potential is imposed by the ac Stark shift of a laser field and can be created for the example of benzene by superposing a hexagonal optical lattice $V_0 \Vh(x,y)$ \cite{SoltanPanahi2011} with a tight dipole trap $\Vdip(x,y)$ as shown in \Fig{Fig1}a (see Appendix). A similar potential can, e.g., also be created by adding a triangular superlattice to the honeycomb lattice (a "benzene lattice" \cite{Supp}) or by using a spatial light modulator, which renders the possibility of almost arbitrary two-dimensional potentials \cite{Fukuhara2013, Preiss2014, Nogrette2014}. In the orthogonal direction a one-dimensional lattice $\Vz$ with depth $V_{0z}$ or a light sheet is applied. A controlled number of fermionic atoms (e.g. $^6$Li atoms with spin states $\pm1/2$) can be loaded into this optical potential \cite{Serwane2011} (see Appendix).  

The molecular orbitals of the artificial molecule are computed using the plane-wave expansion method for vanishing interaction (see Appendix for technical details and parameters). The resulting single-particle orbitals are shown in \Fig{Fig1}b. The six lowest $s$ orbitals represent an energetically separated shell within the inner "C" ring, whereas the higher orbitals (7-18, 25-30) are $sp^2$-hybridized. The orbitals (7-11) and the antibonding orbital (13) form $p$ orbitals pointing along the C ring. The outward-pointing $p$ orbitals (12) and (14-18) lower their energy by maximizing the overlap to the $s$ orbitals of the H atoms. The hybridized orbitals (25-30) show a nodal plane between H and C ring and thus have a higher energy. Energetically in between lie the delocalized $p_z$ orbitals (19-24) with a nodal plane at $z=0$. Benzene is one of the prime examples for this type of delocalized conjugation of a $\pi$ electron system stabilizing the planar ring structure. This property is commonly referred to as aromaticity. Accounting for in total 42 electrons, in benzene three of the six $p_z$ orbitals are occupied. We can tune the energetic order of the orbitals, by tuning the parameters of the potential, e.g. strength and width of the dipole trap as well as the lattice depths $V_0$ and $V_{0z}$ (see \Fig{Fig3}a). 

\textheight 22.8cm

\section{II. Imaging of molecular wave functions}
Quantum gas microscopes have demonstrated optical imaging with single-site resolution in an optical lattice \cite{Bakr2009,Sherson2010}. However, the imaging of molecular wave functions of artificially created molecules requires an optical resolution several times better than the optical lattice spacing. The idea is to overcome the limited spatial resolution by imaging the particles in momentum space. In ultracold atom experiments, this can be achieved by turning off the optical potential, letting the particles freely expand and imaging the particle density $\rho_\text{TOF}(\mathbf{r})$ after a certain time of flight. In the language of molecular physics, this would be equivalent to a sudden removal of all nuclei. During the expansion, the interactions quickly become negligible, because there are only few particles with high momenta. This allows for a free expansion in two dimensions with an unaltered lattice in the $z$ direction avoiding problems with a limited depth of focus of the imaging system. In addition, the interaction among the particles can be tuned to zero via a Feshbach resonance during the expansion. For suitably long expansion times $t$ (see \Supp), the momentum density can be expressed as $\rho_p(\q)\propto\rho_\text{TOF}(\mathbf{p} t /m)$  with momenta $\mathbf{p} =h \q /a$, $a$ the lattice constant of the honeycomb lattice,  and $h$ the Planck constant. 

\begin{figure}
\centering\includegraphics[width=\linewidth]{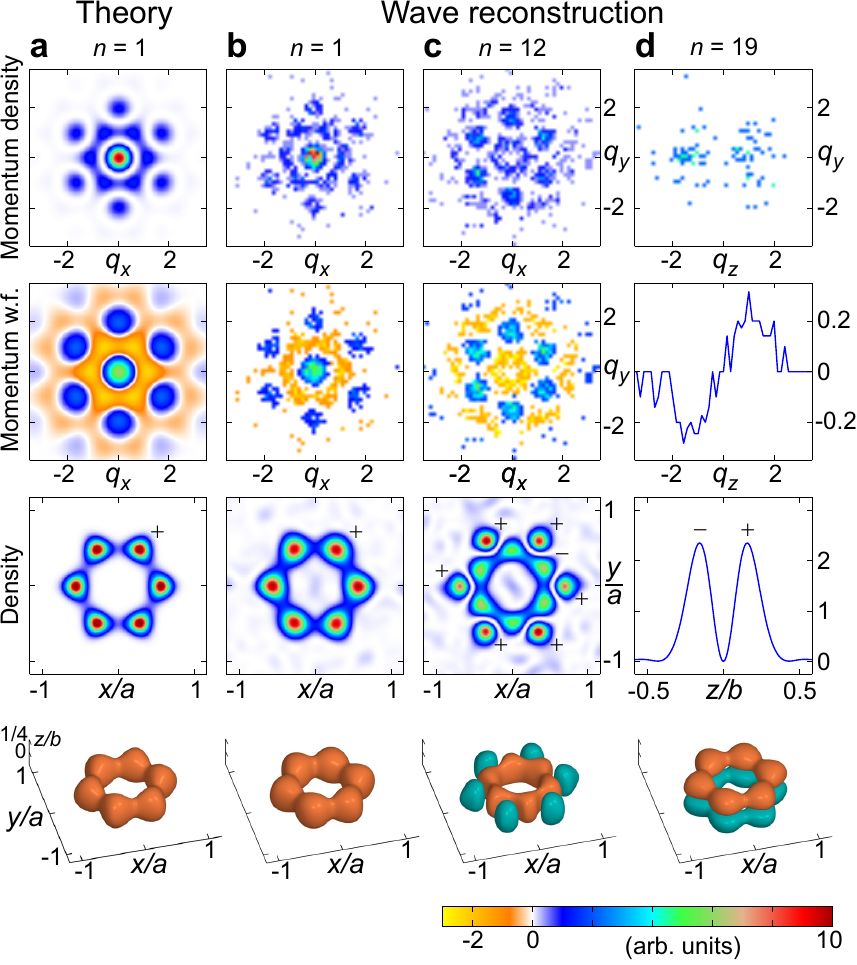}
\caption{Imaging of molecular wave functions via momentum-space mapping. The first row shows the momentum density in two dimensions, the second row the momentum wave function after phase retrieval, the third row the density (the sign of the wave function is indicated), and the fourth row the isosurface of the three-dimensional wave function for $\pm 0.45/a\sqrt{b}$. (a) Theoretical calculation for the lowest orbital ($n\:1$) of the artificial benzene molecule. (b) Simulated time-of-flight measurements with $1000$ observed momenta in the $x$-$y$ plane (randomly distributed) for the orbitals $n\:1$ and (c) $n\:12$. (d) The momentum density in $y$-$z$ direction with $100$ random momenta for the lowest $p_z$ orbital ($n\:19$), integrated momentum wave function and density in $z$ direction. In combination with the reconstructed wave function in $x$-$y$ coordinates (see (b)), this allows one to retrieve the three-dimensional wave function for the lowest $p_z$ orbital (analogous for (b) and (c) using the lowest Wannier orbital in $z$ direction). } \label{Fig2}
\end{figure}

In the example shown in \Fig{Fig2}b, we assume that we can image the momentum density after the time of flight with $P^2=49\!\times\!49$ discrete momenta using a so-called pinning lattice for imaging (see Appendix). Initially preparing  one particle (or two with opposite spin without interaction) only the lowest orbital $n=1$ is occupied and per experimental sequence only $1$-$2$ momenta are recorded. To show the feasibility of the proposed imaging method, we use a random number generator with an accumulated number of $1000$ momenta. Note that in principle satisfying results with fewer observed momenta are possible (see \Supp, Fig.~S4). Figure \ref{Fig2}b (first row) shows that the statics is clearly sufficient to map out the momentum distribution of the orbitals. The phase of the distribution can be retrieved by identifying the nodal lines of the momentum density ($\rho_p(\q)\:0$) and mapping it piecewise to the momentum wave function with $p(\q)=\pm\sqrt{\rho_p(\q)}$ (second row). Since the potential is symmetric, $p(\q)$ can be assumed to be real. The components of $p(\q)$ are the Fourier coefficients of the wave function and therefore allow the direct transformation to orbital wave functions in real space via
\begin{equation}
	\psi(x,y) = \sum_\q p(\q) \ket{\q}
\end{equation}    
with plane waves
\begin{equation}
\ket{\q}=\frac{1}{\Ns}\, \e^{2\pi\I (q_x x + q_y y)/a},   
\label{planeWave}
\end{equation}
lattice constant $a$ and the number of grid points $\Ns$ per reciprocal lattice vector. The resulting density is plotted in the third row of \Fig{Fig2}b. The figure demonstrates that the effective resolution of the imaging method is very high. In principle, the resolution is given by the diameter in momentum space. However, since the momentum wave function drops exponentially to zero for higher values of $\q$, we can assume $p(\q)\approx 0$ for $|\q| \gtrsim 3.5$. Since the diameter in momentum space is not limiting, the spatial resolution in the experiment is mainly influenced by the statistics of the momentum density (see \Supp, Fig.~S4). In contrary, the maximal diameter of the artificial molecule to be imaged is given by the resolution in momentum space and estimated as $P/(2q_x^\text{max}-2q_x^\text{min})$ using the Nyquist theorem ($3.5a$ for our example). 

%%%%%%%%%%%%%%%%%%%%%%%%%%%%%%
%%%%%%%%%%%%%%%%%%%%%%%%%%%%%%

In analogy, higher single-particle orbital wave functions can be imaged as depicted in \Fig{Fig2}c. The preparation of higher orbitals is analogous to the preparation of higher bands in optical lattices via Bragg transfer \cite{Clement2009,Ernst2009} or via amplitude modulation \cite{Heinze2011}. Alternatively, increasing the number of particles in the artificial molecule automatically occupies higher orbitals. In this case, the recorded images correspond to the sum of the respective momentum densities (for vanishing interactions). Imaging the wave functions in three dimensions requires measuring the momentum distribution also in the $z$ direction, where \Fig{Fig2}d depicts the reconstruction of the $p_z$ orbital. Here, a statistics of $100$ particles is sufficient to resolve the orbital structure. In our case, the separable densities in the $x$-$y$ and $z$ direction can be simply combined to the three-dimensional orbital wave functions shown in \Fig{Fig2} (fourth row). 

\section{III. Quantum simulation of\\the many-body problem}
\begin{figure}
\centering\includegraphics[width=\linewidth]{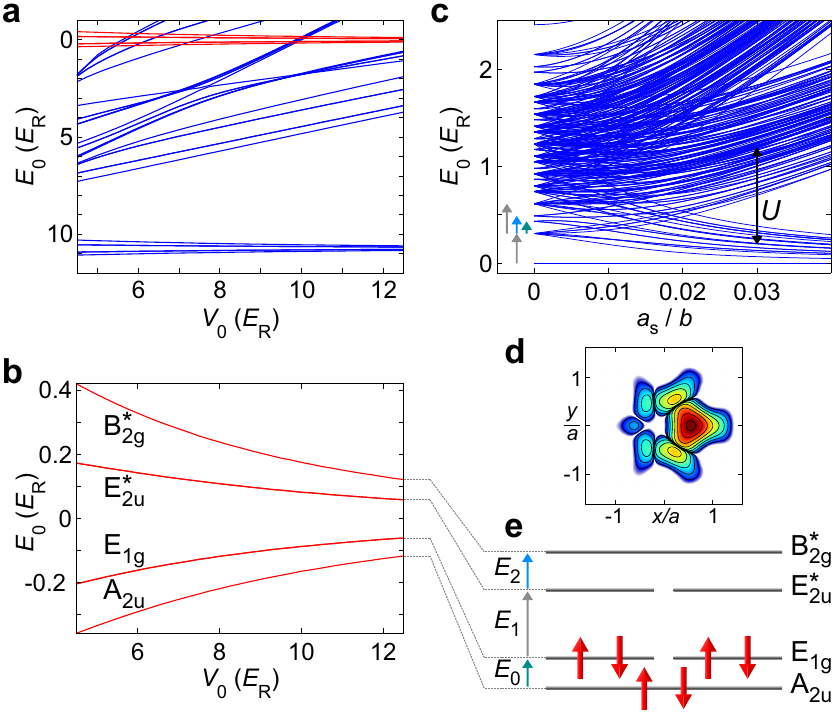}
\caption{Interactions in the conjugated $\pi$ system. (a) Single-particle energy spectrum showing the lowest orbitals ($p_z$ orbitals in red) as a function of the lattice depth $V_0$ (taking the role of the effective nuclei distance $R_0$). See Appendix for parameters. The energy of the $p_z$ orbitals can be tuned independently by lattice depth $V_{0z}$ (here $35\ER$). (b) Close-up showing only the $p_z$ orbitals with three bonding and three antibonding orbitals, where the orbitals 2-3 and 4-5 are degenerate. (c) Energy spectrum of the conjugated system incorporating the on-site interaction $U$ for $V_0=6\ER$ (see Appendix). The colored arrows indicate the lowest excitations (labeled in (e)) for vanishing interaction. (d) Logarithmic density plot of the two-dimensional Wannier function for the $p_z$ (or $s$) band. Each contour indicates a density drop by a factor $1/10$. (e) Level scheme depicting orbitals and spin configuration.  } \label{Fig3}
\end{figure}

The single-particle energies of the orbitals are plotted in Figs.~\ref{Fig3}a and b, where the six $p_z$ orbitals are indicated in red. Because of the separability of the potential in the $z$ direction, these orbitals are analogous to the six lowest $s$ orbitals but with a nodal plane for $z=0$. The lowest and highest orbital are symmetric (A$_\text{2u}$) and antisymmetric (B$^*_\text{2g}$) under rotation (compare Fig. 1b), whereas the second-third and fourth-fifth orbitals are doubly degenerate and labeled as E$_\text{1g}$ and E$^*_\text{2u}$. The lower three orbitals have a bonding character and the higher three orbitals an antibonding character. While molecular potential curves are usually plotted against the nuclei distance $R_0$, the distances are fixed in an optical lattice potential. For the same reason the repulsion of the nuclei, dominating the potential curves at $R_0\to 0$, is not present in optical potentials. However, even though the positions are fixed in our case, we can effectively change the intramolecular distances by tuning the depth $V_0$ of the lattice potential. A deeper lattice potential decreases the density between the sites and is analogous to a larger spacing $R_0$ between the nuclei coordinates.

%%%%%%%%%%%%%%%%%%%%%%%%%%%%%%
%%%%%%%%%%%%%%%%%%%%%%%%%%%%%%

A quantum simulator for molecular problems is needed for non-vanishing interactions, since solving the many-particle problem using the full configuration-interaction method is not feasible for complex molecules even accounting for the molecular symmetries. Unlike electrons, neutral atoms do not interact via Coulomb force, but by either a short-range contact potential or a long-range dipolar potential \cite{Lahaye2009}. For ultracold atoms, the interaction strength between ultracold atoms can be tuned in a wide range via an external magnetic field due to the existence of Feshbach resonances. 
As a concrete example, the $s$-wave scattering length $a_s$ for $^6$Li atoms can be tuned between $-0.2b$ and $0.4b$, where $b=\lambda/2\:532\,\mathrm{nm}$ is the lattice spacing in the $z$ direction. Experimentally, we can probe the energies of the interacting problem (benzene has 42 electrons) by exciting a particle to a reference orbital using a microwave transition, which also changes the internal state. This avoids problems with postinteraction, assuming that the particles in the two different internal atomic states do not interact. Furthermore, the population in this internal atomic state serves as an observable measuring the transfer rate as a function of excitation energy.

To demonstrate how interactions change the electronic configuration, let us turn to the conjugated $\pi$ system, i.e.~the $p_z$ orbitals with delocalized particles, and neglect all interactions with the lower-lying orbitals. When restricting ourselves to this subsystem, the problem can be solved using the full configuration-interaction method. Experimentally, this subsystem is also accessible by populating only the six lowest $s$ orbitals that are equivalent to the $p_z$ orbitals in the $x$ and $y$ direction. In the case of $p_z$ orbitals, the experimental system would incorporate the interaction with the energetically lower-lying particles constituting a much more complex situation. For the description of strongly correlated electron systems, it is often desirable to switch to Wannier orbitals (see Appendix) that are localized at individual sites (potential minima). The Wannier orbital shown in \Fig{Fig3}d allows the formulation of a tight-binding Hubbard Hamiltonian taking into account the tunneling element $t_1 $ and the on-site interaction $U$, where $U$ scales linearly with the scattering length $a_s$. The next-nearest-neighbor tunneling $t_2$ is considerably smaller than $t_1$ but large enough to cause an asymmetry in the $p_z$ band ($E_2$$\neq$$E_0$ in \Fig{Fig3}e). The fermionic many-particle Hamiltonian reads
\begin{equation}
	\hat H= - \sum_{i,\sigma,d}  t_d\ \hat c_{i,\sigma}^\dagger \hat c_{[i+d],\sigma} + \text{c.c.} + U \sum_i \hat n_{i,\up} \hat n_{i,\down},  
\end{equation}
where the operator $\hat c_{i,\sigma}$ annihilates a particle on site $i$ in spin state $\sigma\in \{ \up , \down \}$ and $\hat n_{i,\sigma}=\hat c_{i,\sigma}^\dagger \hat c_{i,\sigma}$ corresponds to the respective particle number. Here, the tunneling matrix elements $t_1$ ($d\:1$) and $t_2$ ($d\:2$) are included and $[i]={i\!\!\mod\!6}$ incorporates the  periodic boundary conditions. 

For vanishing interaction ($a_s\:0$), the energy spectrum (\Fig{Fig3}c) represents all possible single-particle excitations of the half filled band with delocalized particles (\Fig{Fig3}e). Increasing the interaction causes two drastic changes on the "electronic" structure. First, the particles undergo the transition from delocalized orbitals to the Mott insulator state, where the particles occupy the localized Wannier orbitals. At the same time, an interaction gap is formed separating the many-particle bands by the on-site interaction energy $U$. Second, the spins of the particles align in an alternating order on the Wannier orbitals (antiferromagnetic alignment), thereby gaining the second-order energy $2t_1^2/U$ per particle. The lowest Mott band contains possible spin excitations from this antiferromagnetic ground state. For the interaction strength $a_s=0.02b$, the ratio $U/t_1\approx 4$ is in the vicinity of the proposed values for aromatic hydrocarbons which are still under debate \cite{Parr1950,Verges2010}. The tunability of interaction strength and potential depth allows an independent control of the ratios $U/t_1$ and $t_2/t_1$ (or the next-neighbor interaction for long-rage interactions \cite{Lahaye2009}). 
 
\section{IV. Correlated electron dynamics}
\begin{figure}
\centering\includegraphics[width=\linewidth]{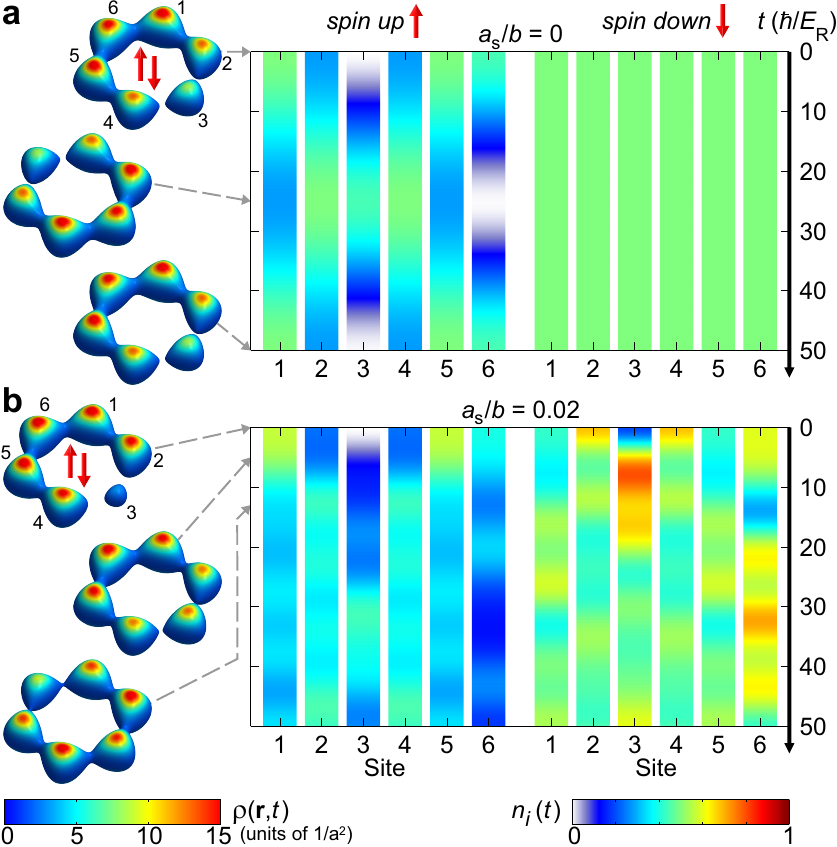}
\caption{Correlated dynamics after particle removal. (a) Quantum revival for the noninteracting system and (b) dynamics in the antiferromagnetically correlated state for $a_s/b\:0.02$ and $V_0=6\ER$. The plots display the occupation number of the Wannier state at site $i$ of the ring for spin up (left) and spin down (right) as a function of time $t$. The images on the left depict the summed density of both spin states. The three-dimensional isosurface is plotted for $6/a^2b$, where the coloring refers to the two-dimensional density.} \label{Fig4}
\end{figure}

One of the outstanding advantages of ultracold atomic systems is that the dynamical time scale is on the order of milli- or microseconds. Therefore, artificial molecules would allow monitoring of dynamical processes taking place on femto- or attoseconds time scales in real molecules. As an example, we compute the hole dynamics after the removal of a particle (see \Fig{Fig4}). Again, we restrict ourselves to the conjugated $\pi$ system, for which the dynamics can be computed. The site-selective manipulation has been demonstrated for an optical lattice using a tightly focused addressing beam \cite{Weitenberg2011,Fukuhara2013}. After an evolution time $t$, suddenly ramping up the optical lattice allows freezing the particle number distribution at the individual sites of the ring. Using fluorescence imaging, we can count the occupation number $n_{i,\up} +n_{i,\down}$ on each site of the artificial molecule \cite{Bakr2009,Sherson2010}. The spin state can be identified by selectively removing one of the spin states before imaging via resonant light \cite{Weitenberg2011}. 

In \Fig{Fig4} the site-resolved dynamics is plotted after the removal of a spin-up particle at site $i\:3$. Without interactions (\Fig{Fig4}a), the spin-down particles are not influenced by the particle removal, whereas the initial hole in the spin up component oscillates between sites $i\:3$ and $6$. This quantum revival occurs at the revival time $T=h/E_0$ (with $E_0\:0.13\ER$). This corresponds to the energy difference between A$_\text{2u}$ and E$_\text{1g}$ orbitals. For $^6$Li, the revival time corresponds to $0.27\mathrm{ms}$ for a recoil energy $\ER\:h \!\times\! 29.4\mathrm{kHz}$ at a lattice wavelength $\lambda=1064\,$nm. Switching on the (repulsive) interaction among the particles (\Fig{Fig4}b), the dynamical behavior becomes very complex. For $t\:0$, the antiferromagnetic order is apparent in the spin-down component causing the almost vanishing total density on site $i\:3$. The neighboring sites are highly occupied with the down component, which fills the empty site on short time scales. This interferes strongly with the dynamics of the up component leading to complex density modulations. \textit{A priori}, it is not clear if the spin order persists at the typical temperatures of ultracold experiment. Therefore, the dynamics shown in \Fig{Fig4} incorporates a finite temperature of $0.05\ER/k_\mathrm{B}$ with $k_\mathrm{B}$ the Boltzmann constant, which corresponds to $70\,\mathrm{nK}$ for $^6$Li, demonstrating that the quantum simulation of electronic dynamics is feasible with state-of-the-art temperatures. 

%%%%%%%%%%%%%%%%%%%%%%%%%%%%%%
%%%%%%%%%%%%%%%%%%%%%%%%%%%%%% 
 
\section{V. Conclusions}
In conclusion, ultracold atoms have great potential as a quantum simulator for molecular physics, where solving the full many-particle problem is still out of reach using either classical or quantum computers \cite{Wecker2014}. This includes the formation of molecular orbitals, electronic interactions as well as the time-resolved monitoring of a broad range of dynamical processes. With the proposed quantum emulation, we hope to shed light on open questions regarding the influence of electronic correlations and localization in molecular systems. In particular, the ability of directly observing spatial correlations goes far beyond the available techniques in molecular physics, where correlations are probed indirectly by photoionization processes. This also touches on the fundamental questions of how interactions change the picture of effective molecular single-particle orbitals. The electronic  correlations can fundamentally affect the electron dynamics in a molecular system. While we discuss the creation of an artificial benzene molecule using state-of-the-art experimental techniques, the results can be easily generalized covering more complex molecules. In particular, spatial light modulators can be used to create arbitrary two-dimensional molecular structures \cite{Nogrette2014}. To emulate nonplanar molecules, one could utilize three-dimensional nonseparable lattice potentials formed by several interfering laser beams \cite{Petsas1994} or by holographic projection techniques. The nuclei positions, kept fixed in the present study, can be controlled using slowly varying time-dependent optical potentials (e.g. using a spatial light modulator or an accordion lattice \cite{Al-Assam2010}) or coupled to phonons using other atomic species such as ions \cite{Bissbort2013}. This allows simulation of the electronic response to vibrational modes. In principle, one could also imagine accessing regimes where the Born-Oppenheimer approximation breaks down. 

\section{Acknowledgments}
We acknowledge funding by the Deutsche Forschungsgemeinschaft (Grants SFB~925 and the The Hamburg Centre for Ultrafast Imaging (CUI)). 

\appendix
\section{Appendix A: Proposed experimental sequence}
The optical potential for the artificial benzene molecule is generated by three laser beams in the $x$-$y$ plane, which create a honeycomb hexagonal optical lattice $V_0 \Vh(x,y)$, and an orthogonal one-dimensional lattice $V_{0z} \cos^2(\pi z / b)$ (see \Supp). Superposing this lattice potential with a tight dipole trap $\Vdip(x,y)=- \Vd V_0 \e^{ - 2 (x^2+y^2 ) / \wdip^2 }$ centered on a plaquette of the lattice as sketched in \Fig{Fig1}a causes a finite-size system with a hexagonal inner and outer ring, where $V_0$ is the height of the potential (in units of the recoil energy $\ER=h^2  / 2 \lambda^2 m$ with wavelength $\lambda$ and atomic mass $m$). For concreteness, we choose if not stated otherwise the lattice depths $V_0\:11\ER$ and $V_{0z}=35\ER$, the relative strength $\Vd\:10$ of the dipole potential and the width $\wdip\:4a$ of the dipole trap. Here, $a\:2\lambda/3$ is the width of the unit cell in the honeycomb lattice, whereas $b\:\lambda/2$ is the lattice spacing in the $z$ direction. The orbital order can be tuned by adjusting the lattice depths $V_0$ and $V_{0z}$ (see \Fig{Fig3}a).

A controlled number of ultracold fermionic atoms can be prepared in the tight two-dimensional potential formed by a dipole trap and a single antinode of the orthogonal lattice. Atoms above a certain energy are expelled from the trap via a strong magnetic field gradient \cite{Serwane2011}. When adiabatically ramping up the honeycomb lattice, the particles fill up the lowest orbitals of the artificial molecule. To image the momentum distribution a time-of-flight measurement is performed, where the atoms expand for a certain time. Subsequently, a deep so-called pinning lattice is ramped up that holds the atoms in place while they are illuminated by an optical molasses \cite{Bakr2009,Sherson2010}. The pinning lattice can be a square lattice of the same wavelength $\lambda$ containing $P^2$ central lattice sites. Using a high-resolution imaging system the individual sites of the pinning lattice can be resolved \cite{Bakr2009,Sherson2010}. In addition to the observed momenta, the number of particles is also retrieved.

For observing the momentum density in the $z$ direction, a three-dimensional expansion can be performed. In this case, the imaging (from the $x$ direction) needs a large depth of field and is therefore restricted to a larger optical resolution. However, for dilute filling of the optical lattice, the atomic positions can still be identified even if the optical resolution is several times the lattice spacing \cite{Karski2009} (see \Supp). 

\section{Appendix B: Computation of orbitals}
We compute the molecular orbitals by applying the plane-wave expansion method for nonperiodic systems. By decomposing the two-dimensional potential in Fourier components $V(x,y)=\sum_\q v_\q \ket{\q}$, the Hamiltonian matrix can be calculated in the plane-wave basis \eqref{planeWave}. Here, $\Ns$ is the number of unit cells per dimension, $\q\:(q_x,q_y)$ and $q_i\:-1/2$$+$$\nu/\Ns$$+$$d$ are the wave vectors with $d\:\{-D,-D$$+$$1,...,D\}$ and $\nu\:0,1,...,\Ns$$-$$1$ ($\nu\:1/2,...,\Ns$$-$$1/2$ for $\Ns$ odd). The eigenvectors of the Hamiltonian matrix of dimension $(2D+1)^4 \Ns^4$ represent the solution for the single-particle orbitals in the plane-wave basis  $\ket{\q}$. The orthogonal optical lattice in the $z$ direction separates. Assuming that only the central lattice plane is filled in the experiment, we can restrict ourselves to the Wannier function $w_z^{(n)}(z)$ of band $n$ in the $z$ direction. For the orbitals in Figs.~\ref{Fig1} and \ref{Fig3} we use $\Ns\:5$, $D\:7$ and $\Ns\:15$, $D\:5$ for \Fig{Fig2}a. 

\section{Appendix C: Molecular Wannier functions}
The Wannier orbitals at site $i$ of the C ring can be constructed from the six $s$ (or $p_z$) orbitals $\ket{\psi^{(s)}_j}$ via the basis transformation $\ket{w_i}=\sum_j c_{ij}\,\ket{\psi^{(s)}_j}$ with suitable coefficients $c_{ij}$  (see \Supp~and Ref.~\cite{Luhmann2014}). For the lattice depths $V_0\:6\ER$ and $V_{0z}\:25\ER$ ($\Vd\:10$, $\wdip\:4a$), the tunneling element is $t_1\:0.154\ER$, the next-nearest-neighbor tunneling $t_2\:-0.0094\ER$ and the on-site interaction $U\:32.9 \ER a_s /{b}$ ($U_{(p_z)}\:22.7 \ER a_s /{b}$ for the $p_z$ orbital), where $a_s$ is the $s$-wave scattering length and $b=\lambda/2$ (cf. \Fig{Fig3}c, d and \Fig{Fig4}). 

%\bibliography{references}

%merlin.mbs apsrev4-1.bst 2010-07-25 4.21a (PWD, AO, DPC) hacked
%Control: key (0)
%Control: author (0) dotless jnrlst
%Control: editor formatted (1) identically to author
%Control: production of article title (0) allowed
%Control: page (1) range
%Control: year (0) verbatim
%Control: production of eprint (0) enabled
%

\end{document}